\documentclass[12pt]{article}
\usepackage{graphicx}
\usepackage{dcolumn}
\usepackage{bm}
\usepackage{latexsym}

\begin{document}
\thispagestyle{empty}

\title{Useful equation of  tridiagonal matrices in application to electron transport through a quantum wire}

\maketitle

\begin{center}
\author{T. Kwapi\'nski\footnote[1]{e-mail address:
tomasz.kwapinski@umcs.lublin.pl}\\}

\vspace{1cm}

Institute of Physics, M. Curie-Sk\l odowska University \\
Pl. M. Curie-Sk\l odowskiej 1, 20-031 Lublin, Poland
\end{center}



\begin{abstract}
In this paper the transmittance through a quantum wire connected
with two electron reservoirs is calculated and non-trivial
transformation between the evolution operator method and the
Green's function technique is reported. To show this equivalence
an analytical nonlinear formula which concerns symmetrical
tridiagonal matrices is proofed. This formula  connects the
cofactor and three determinants of tridiagonal matrices.
\end{abstract}


\section{\label{sec10}Introduction}
There are many mathematical, e.g. \cite{Meu,Kam,Fort}, and
physical, e.g. \cite{Fort,Mar,Kwa1,Kwa2}, problems which are
described by means of tridiagonal matrices. For such matrices many
analytical properties were reported especially to find the
determinant or the inverse of this kind of matrices, cf.
Refs.~\cite{Fort,Schle,Hu,Sim,Fon}. These properties allows us to
derive analytically many physical and mathematical problems. In
this work we report one interesting equation which concerns
symmetric tridiagonal matrices. This equation can be useful in
obtaining solutions of many mathematical and physical problems
with tridiagonal matrices and finally it can lead to analytical
expressions for physical quantities or in  mathematical
calculations.

As an example of this equation we obtain the transmittance through
a quantum wire between two eternal electrodes. Electronic
properties of quantum wires are very important mainly from the
practical point of view i.e. due to their potential applications
in nanoelectronic - they are the thinnest possible electric
conductors.   We obtain the transmittance through a wire by means
of the evolution operator (EO) method and show that the final
analytical relation for the transmittance is the same as that one
obtained within the Green's function (GF) technique. Note, that
there is no trivial transformation between the final results
obtained within these methods. Until now, analytical derivation of
the quantum wire transmittance, obtained by means of the evolution
operator method, was not reported in the literature.

\section{\label{sec200}Nonlinear  relation on tridiagonal matrixes}
Lets consider tridiagonal symmetrical square matrix $\textbf{A}_N$
(dimension $N\times N$) written in the following form:
\begin{eqnarray}
\textbf{A}_N=\left(%
\begin{array}{cccccc}
  \alpha & \beta & 0 & 0 & \cdots & 0 \\
  \beta & \alpha & \beta & 0 &  &  \\
  0 & \beta & \alpha & \beta &  &  \\
  0 & 0 & \beta & \alpha &  &  \\
   \vdots &  &  &   & \ddots  &   \\
  0 &  &  &  &  & \alpha \\
\end{array}%
\right)_{N\times N}
\end{eqnarray}
where $\alpha$ and $\beta$ are real numbers. For this matrix the
following nonlinear equation is satisfied:
\begin{equation}\label{main}
\left[
\texttt{cof}(\textbf{A}_N)_{N,1}\right]^2={A^2_{N-1}}-{A_{N-2}}{A_{N}}
\end{equation}
where ${A_N}$ is the determinant of  $\textbf{A}_N$ matrix,
$A_N=\det\textbf{A}_N$, (we consider ${A_N}\neq 0$) and
$\texttt{cof}(\textbf{A}_N)_{N,1}$ means the algebraic complement
(cofactor) of  $\textbf{A}_N$ matrix versus the element $(N,1)$ or
$(1,N)$. Note, that Eq.~\ref{main} connects the cofactor of the
matrix $\textbf{A}_N$ and three determinants of this matrix for
different dimensions, i.e. $N, N-1$ and $N-2$. It is interesting
that this relation is connected with the Fibonacci and Lucas
numbers, e.g. \cite{Str,Str2}, where only the integer numbers
appear and, moreover, it can be transformed to a kind of
trigonometrical relation because the determinant of tridiagonal
matrixes can be expressed by means of the second kind Chebyshev
polynomials. Here we show a simple proof of this relation using
only well known properties of tridiagonal matrixes with arbitrary
real $\alpha$ and $\beta$.

\emph{Proof of Eq.~\ref{main}.} The cofactor in Eq.~\ref{main} can
be expressed as follows:
\begin{eqnarray}
 \texttt{cof}(\textbf{A}_N)_{N,1} \equiv B_{N-1}=\det \left(%
\begin{array}{cccccc}
  \beta & 0 & 0 & 0 & \cdots & 0 \\
  \alpha & \beta & 0 & 0 &  &  \\
  \beta & \alpha & \beta & 0 &  &  \\
  0 &  \beta & \alpha & \beta & 0&  \\
   \vdots &  &  &   & \ddots  &   \\
  0 &  &  &  &  & \beta \\
\end{array}%
\right)_{N-1\times N-1} \label{cof}
\end{eqnarray}
Using Eq.~\ref{cof} the left side of Eq.~\ref{main} can be
expressed only by means of $\beta$ elements and the dimension of
the matrix $\textbf{A}_N$, i.e.
\begin{equation}\label{proof1}
\left[
\texttt{cof}(\textbf{A}_N)_{N,1}\right]^2={B^2_{N-1}}=\beta^{2N-2}
\end{equation}
Now, we proof that the right side of Eq.~\ref{main} can be also
expressed in the same form. In our calculation we use the
following recurrence relation, e.g. \cite{Mar,Hu},
\begin{equation}\label{proof2}
{A_N}=\alpha {A_{N-1}}-\beta^2 {A_{N-2}}
\end{equation}
Using this equation the right side of Eq.~\ref{main} can be
written as follows:
\begin{eqnarray}\label{proof3}
&&{A^2_{N-1}}-{A_{N-2}}{A_{N}}= \nonumber\\ && (\alpha {A_{N-2}}-
\beta^2 {A_{N-3}})^2-  {A_{N-2}} (\alpha^2 {A_{N-2}}-\alpha\beta^2
{A_{N-3}}-\beta^2 {A_{N-2}})
\end{eqnarray}
and after some algebra one obtains
\begin{eqnarray}\label{proof3b}
\beta^2( {A^2_{N-2}}-{A_{N-3}}{A_{N-1}})
\end{eqnarray}
which is also the recurrence relation and in the end it leads to
the final result for the right side of Eq.~\ref{main}:
\begin{eqnarray}\label{proof4}
\beta^{2(N-2)} ({A^2_{1}}-{A_{0}}{A_{2}})=\beta^{2N-2}
\end{eqnarray}
Here we use the explicate forms of $\textbf{A}_N$ determinants,
i.e. ${A_{2}}=\alpha^2-\beta^2$, ${A_{1}}=\alpha$ and ${A_{0}}=1$.
The last result, Eq.~\ref{proof4}, is the same one as
Eq.~\ref{proof1} (obtained for the left side of Eq.~\ref{main}).
Thus the main equation of this paper, Eq.~\ref{main},  has been
proofed.\rule{5pt}{5pt}

\section{\label{sec20}Electron transport through a quantum wire}

In this section we consider the electron transport through a
quantum wire coupled with the left and right electrodes. The wire
is represented by a finite, straight chain of $N$ sites with the
nearest-neighbour hoppings, i.e. electrons cannot tunnel between
the next-neighboour sites. For simplicity we consider the case of
no electron-electron correlations. The Hamiltonian of considered
system, written in the second quantization notation, is given by
$H = H_0 + V$, where
\begin{equation}
H_0 = \sum_{\vec k\alpha=L,R} \varepsilon_{\vec k\alpha} c^+_{\vec
k\alpha} c_{\vec k\alpha} +\sum^N_{i=1} \varepsilon_0 c^+_{i}
c_{i}\,, \label{eq1}
\end{equation}
is a single electron Hamiltonian which stands for on-site energies
and
\begin{equation}
V = \sum_{\vec kL} V_{\vec kL} c^+_{\vec kL}c_1 +
    \sum_{\vec kR} V_{\vec kR} c^+_{\vec kR} c_N +
    \sum^{N-1}_{i=1} V_{N} c^+_{i} c_{i+1} + {\rm h.c.}\,
\label{eq2}
\end{equation}
represents the interaction energy. Here the operators $c_{\vec
k\alpha}(c^+_{\vec k\alpha})$, $c_i(c^+_i)$  are the annihilation
(creation) operators of the electron in the lead $\alpha$ ($\alpha
= L, R$) and at site $i$ in the wire, respectively. The wire sites
are described by the electron energy levels, $\varepsilon_0$,
which are the same for all sites. Our wire is coupled to both
electrodes through the tunneling barriers with the transfer-matrix
elements $V_{\vec kL}$ (the first wire site: $i=1$) and $V_{\vec
kR}$ (the last wire site: $i=N$). $V_{N}$ denotes the hopping
integrals between the nearest-neighbour wire sites and are the
same for all sites. In the above Hamiltonian $\varepsilon_{\vec
kL/R}$ corresponds to the energy of electrons with the wave vector
$\vec k$ in the left or right electrodes.

\subsection{\label{sec21}Green's function method}
For considered here system the transmittance,  $T(\varepsilon)$,
can be expressed by means of the retarded Green's function between
the first and the last quantum wire sites,
$G^r_{1N}(\varepsilon)$, cf. \cite{Mar,Kwa1,Kwa2}:
\begin{eqnarray}\label{tt1}
T(\varepsilon)=\Gamma^2 |G^r_{1N}(\varepsilon)|^2
\end{eqnarray}
Using equation of motion for the retarded Green's function and the
Hamiltonian of the system one can find the following relation for
the element $G^r_{1N}(\varepsilon)$:
\begin{eqnarray}\label{gg1}
G^r_{1N}(\varepsilon)={\texttt{cof}(\textbf{C}_N)_{N,1} \over
{C}_N}
\end{eqnarray}
where  $\textbf{C}_N$ is a kind of symmetric tridiagonal matrix
($C_N$ is its determinant) and can be written in the form
\begin{eqnarray}\label{matrix}
\textbf{C}_N=\left(%
\begin{array}{cccccc}
  \varepsilon_0-\varepsilon_{\vec kL}+i{\Gamma\over 2} & -V_N & 0  & \cdots & 0 \\
  -V_N & \varepsilon_0-\varepsilon_{\vec kL} & -V_N  &  &  \\
  0 & -V_N & \varepsilon_0-\varepsilon_{\vec kL} &   &  \\
   \vdots &  &   & \ddots  &   \\
  0 &  &  &   & \varepsilon_0-\varepsilon_{\vec kL}+i{\Gamma\over 2} \\
\end{array}%
\right)_{N\times N}
\end{eqnarray}
Here we use  the wide-band limit approximation i.e.
$\Gamma=\Gamma^{L/R}=2\pi \sum_{\vec kL}V_{\vec kL} V^*_{\vec
kL}\delta(\varepsilon-\varepsilon_{\vec kL})=2\pi |V_L|^2/D$ where
we assume that $V_{\vec kL}$ elements do not depend on the wave
vector $\vec k\alpha$ ($V_{\vec k \alpha}=V_{\alpha}$) and $D$ is
the effective band width of the left electrode.  Using
Eq.~\ref{gg1} the transmittance can be written as follows:
\begin{eqnarray}\label{tt2}
T(E)=\Gamma^2 {|\texttt{cof}(\textbf{C}_N)_{N,1}|^2 \over
|{C}_N|^2}
\end{eqnarray}
It is worth noting that the determinant of the matrix
$\textbf{C}_N$ as well as the cofactor can be obtained
analytically which leads to the analytical formula for the
transmittance, see e.g. \cite{Mar}.

\subsection{\label{sec22}Evolution operator technique}
In this subsection we obtain the transmittance through the wire
using the evolution operator method. The current flowing through
the system (or the transmittance) can be expressed by means of
appropriate evolution operator matrix elements, e.g.
\cite{Tar,Kwa3}, i.e.
\begin{eqnarray}
j_L(t) = -e{dn_L(t)\over dt}=  -e{d \over dt}\sum_{\vec kL}
n_{\vec kL}(t) = \sum_{\vec kL} \sum_{\beta}n_{\beta}(t_0)|U_{\vec
kL,\beta}(t,t_0)|^2 \,, \,
 \label{curr1}
\end{eqnarray}
where $n_L(t)$ means the electron occupation of the left electrode
at time $t$, $n_{\beta}(t_0)$ represents the initial filling of
the corresponding single-particle states ($\beta=i,\vec kL, \vec
kR$) and $U(t,t_0)$ is the evolution operator matrix element which
satisfies the following equation of motion (in the interaction
representation, $\hbar=1$), cf. \cite{Tar,Gri,Tar2}:
\begin{equation}
 i{\partial\over\partial t} U(t,t_0) = \tilde V(t)\,U(t,t_0)
\label{eq3}
\end{equation}
Here $\tilde V(t) = U_0(t,t_0) \, V(t) \, U^+_0(t,t_0) $ and
$U_0(t,t_0) = T\exp\left(i\int^t_{t_0} dt' H_0(t')\right)$.
Assuming the wide band limit approximation the current,
Eq.~\ref{curr1},  can be written in the following form:
\begin{eqnarray}
j_L(t)=-\Gamma^L  (\sum_{\vec kL} n_{\vec kL}(0) |U_{1 \vec kL}(t)|^2+\sum_{\vec kR} n_{\vec kR}(0) |U_{1 \vec kR}(t)|^2) &\\
 \vspace{2cm} -2 Im\left(V_L\sum_{\vec kL}  n_{\vec kL}(0) e^{i(\varepsilon_{\vec kL}-\varepsilon_{1})t}
 U_{1 \vec kL}(t)\right) \nonumber\,
\label{curr2}
\end{eqnarray}
To obtain the transmittance, one can symmetrize the average
current flowing through the system: $\langle j(t)\rangle =\langle
j_L(t)\rangle=\left( \langle j_L(t)\rangle-\langle j_R(t) \rangle
\right)/2$ and finally we find the Landauer formula for the
currant:
\begin{eqnarray}
\langle j(t)\rangle = \int d\varepsilon (f_L(\varepsilon) -
f_R(\varepsilon)) T(\varepsilon)
\end{eqnarray}
where the transmittance, $T(\varepsilon)$, is expressed by means
of the evolution operator elements:
\begin{eqnarray}
T(\varepsilon)&=&  {\Gamma^L\over 2D} \left(|\langle U_{N, \vec
kL}(t)\rangle|^2-|\langle U_{1, \vec kL}(t)\rangle|^2\right) \nonumber\\
&-& {\rm Im} {V_L\over D} \langle e^{i
(\varepsilon-\varepsilon_0)t} U_{1, \vec kL}(t) \rangle
\label{eq9}
\end{eqnarray}
These elements satisfy the following set of differential equations
\begin{eqnarray}
{\partial\over\partial t} U_{i, \vec kL}(t) &=&
   -i V_N (U_{i+1, \vec kL}(t)+U_{i-1, \vec kL}(t)) \nonumber\\
&-&\delta_{i, 1} \left( i V_{L} e^{i
(\varepsilon_0-\varepsilon_{\vec kL})t} +
   \Gamma^L U_{1, \vec kL}(t)/2\right) \nonumber\\
   & -&\delta_{i, N} \Gamma^R U_{N, \vec kL}(t)/2\,
\label{diff}
\end{eqnarray}
Note, that there are $N$ complex equations and for $N=1$ or $N=2$
simple analytical solutions for $U_{i, \vec kL}(t)$ exist. In
general, for arbitrary $N$, it can be shown that the above set of
differential equations can be written in the following matrix
notation:
\begin{eqnarray}
\textbf{C}_N \hat{U}=\hat{J} \label{aa}
\end{eqnarray}
where $\textbf{C}_N$ is defined according to Eq.~\ref{matrix},
$\hat{U}$ is $N$-element vector,  $\hat{U}=[U_{1, \vec kL}(t),
\cdots , U_{N, \vec kL}(t)]_N$ and $\hat{J}=[i V_{L} e^{i
(\varepsilon_0-\varepsilon_{\vec kL})t}, 0, \cdots , 0]_N$. The
formal solution for the evolution operator matrix elements reads:
\begin{eqnarray}
U_{i, \vec kL}(t) &=& (\textbf{C}_N)^{-1}_{i,1}  {V_L} e^{i
(\varepsilon_0-\varepsilon_{\vec kL})t} \label{123}
\end{eqnarray}
where $(\textbf{C}_N)^{-1}_{i,1}$ means $(i,1)$ element of the
inverse matrix $(\textbf{C}_N)^{-1}$.  Using the above solution we
find that:
\begin{eqnarray}
\langle|U_{1, \vec kL}(t)|^2\rangle &=& 2 \pi V_L^2 {{|{\hat
{C}_{N-1}}+i{\Gamma \over 2} {\hat C_{N-2}}|^2}\over{|{
C_{N}}|^2}} \label{124}
\end{eqnarray}
\begin{eqnarray}
\langle |U_{N, \vec kL}(t)|^2\rangle &=& 2 \pi V_L^2 {{|
\texttt{cof}(\textbf{C}_N)_{N,1}|^2}\over{|{ C_{N}}|^2}}
\label{125}
\end{eqnarray}
\begin{eqnarray}
-{\rm Im} {V_L\over D} \langle e^{i (\varepsilon-\varepsilon_0)t}
U_{1, \vec kL}(t) \rangle &=& {\Gamma^2 \over 2 |C_N|^2  } \left(
2 \hat{C}_{N-1}^2-\hat{C}_{N-2} \hat{C}_N + {\Gamma^2 \over 4}
\hat{C}_{N-2}^2 \right) \label{126}
\end{eqnarray}
where the tridiagonal and symmetric  matrix $\hat \textbf{C}_N$
corresponds to the matrix $\textbf{C}_N$ for $\Gamma=0$ (wire non
coupled with electrodes) and  $\hat C_N$ is the determinant of
 $\hat \textbf{C}_N$. The above relations allows us
to write the transmittance, Eq.~\ref{eq9}, in the following form:
\begin{eqnarray}
T(\varepsilon)={\Gamma^2 \over 2 |C_N|^2} \left\{ \left|
\texttt{cof}(\textbf{C}_N)_{N,1}\right|^2+{\hat C_{N-1}^2}-{\hat
C_{N-2}}{\hat C_{N}} \right\} \label{127}
\end{eqnarray}
This relation is very important analytical relation obtained by
means of the evolution operator method. Note, that the
transmittance of $N$-site wire is expressed only by the
tridiagonal matrix elements which can be obtained fully
analytically. Note, that analytical calculations for the
transmittance of a quantum wire, obtained by means of the
evolution operator method, were not reported in literature (the
set of differential equations for $U_{i, \vec kL}$ were solved
numerically, e.g. \cite{Kwa3}). In the next subsection we show
that the above relation is equivalent to Eq.~\ref{tt2} obtained by
means of the Green's function technique.

\subsection{\label{sec23}Equivalence between EO and GF methods}

In this part we show that using the special matrix relation which
was introduced in Sec.~\ref{sec200}, Eq.~\ref{main}, we can
transform the transmittance, Eq.~\ref{127} (evolution operator
method), to Eq.~\ref{tt2} (Green's function method). Note, that
Eq.~\ref{127} for the transmittance seems very similar to
Eq.~\ref{tt2} if only
\begin{eqnarray}
{\hat C_{N-1}^2}-{\hat C_{N-2}}{\hat C_{N}} =0.
\end{eqnarray}
But in general, this equivalence does not occur and one should
obtain the appropriate determinants more carefully.

It is worth noting that the matrix $\textbf{C}_N$ has only two
complex elements i.e. $(1,1)$ and $(N,N)$ elements, and the
cofactor, $\texttt{cof}(\textbf{C}_N)_{N,1}$, is real. It means
that the square of the absolute value for this cofactor is the
same as the real square, i.e. $\left|
\texttt{cof}(\textbf{C}_N)_{N,1}\right|^2=\left[
\texttt{cof}(\textbf{C}_N)_{N,1}\right]^2$. Now, using
Eq.~\ref{main} the transmittance, Eq.~\ref{127}, is expressed in
the following form:
\begin{eqnarray}
T(\varepsilon)=\Gamma^2 { [ \texttt{cof}( \textbf{C}_N)_{N,1}]^2
\over |C_N|^2}
 \label{last}
\end{eqnarray}
The above result for the transmittance, Eq.~\ref{last}, is the
same as that one obtained by means of the retarded Green's
function, Eq.~\ref{tt2}.

\section{\label{sec30}Conclusions}

In this paper the analytical formula for the transmittance through
a quantum wire has been obtained by means of the evolution
operator technique, Eq.~\ref{127}. In our calculation the set of
differential equations, Eq.~\ref{diff}, has been resolved and
helpful relation on tridiagonal matrixes has been used,
Eq.~\ref{main}. The transmittance, Eq.~\ref{127} has been
transformed to well known in the literature relation obtained by
means of the retarded Green's function technique, Eq.~\ref{tt2}.
Our analytical calculations are not trivial and can be helpful in
many physical phenomena where tridiagonal matrices appear. It is
worth noting that the evolution operator method can be used also
to describe time-dependent phenomena in nanostructures like e.g.
photon assisted tunnelling in quantum wires \cite{Kwa3}, driven
quantum dot systems \cite{Tar}, chemisorption processes or
collision of atoms with a metallic surface \cite{Gri,Tar2},
decoherence processes in qubits and many others.

\section*{\label{sec50}Acknowledgements}
This work has been supported by  Grant No N N202 1468 33 of the
Polish Ministry of Science and Higher Education.


\newpage
\noindent

\end{document}